# Should Citations be Counted Separately from Each Originating Section?[1]

Articles are cited for different purposes and differentiating between reasons when counting citations may therefore give finer-grained citation count information. Although identifying and aggregating the individual reasons for each citation may be impractical, recording the number of citations that originate from different article sections might illuminate the general reasons behind a citation count (e.g., 110 citations = 10 Introduction citations + 100 Methods citations). To help investigate whether this could be a practical and universal solution, this article compares 19 million citations with DOIs from six different standard sections in 799,055 PubMed Central open access articles across 21 out of 22 fields. There are apparently non-systematic differences between fields in the most citing sections and the extent to which citations from one section overlap with citations from another, with some degree of overlap in most cases. Thus, at a science-wide level, section headings are partly unreliable indicators of citation context, even if they are more standard within individual fields. They may still be used within fields to help identify individual highly cited articles that have had one type of impact, especially methodological (Methods) or context setting (Introduction), but expert judgement is needed to validate the results.

## 1   Introduction

Although citations perform different functions, citation counts reveal nothing about the reasons why an article has been cited. This is unfortunate since roles (e.g., methods, theory) may be differently valued and some citations are perfunctory, for example when introducing concepts (Case & Higgins, 2000; Small, 1978). In addition, highly cited papers are often methods-related, rather than reporting discoveries or advancing theory (Wouters, 1999). An analysis of the 1000 most cited papers in the PMC Open Access collection, for example, found 55% to be cited for their method contribution (90% of the top 100 papers), 60% of which were computational methods (Small, 2018). It has also been suggested that Background/Literature Review citations may be the least central to the citing paper and should therefore be assigned a lower value than other citations (Zhao, Cappello & Johnston, 2017), suggesting that counting citations by source section may give value. It would therefore be useful to differentiate between different types of citation when evaluating the impact of research (Voos & Dagaev, 1976) or when identifying the foundational papers in a field. Since citation context is difficult to assess on a large scale (e.g., see: Jha, Jbara, Qazvinian, & Radev, 2017), a simple method would be to record the sections from which the citations originated. For example, citation counts could be broken into six parts: the number originating from the Introduction, Background, Methods, Discussion, Results and Conclusion sections. This seems to have become technically possible on a large scale due to electronic publishing, the use of standard XML formats for research articles (NCBI, 2015) and a shift towards open access publishing. Nevertheless, there is no clear demonstration yet that counting citations by section is technically possible, would give different solutions from standard citation counts, and whether there are important disciplinary differences to consider. Whilst these seem likely to be true, it is important to demonstrate and investigate them to put the solution on record and reveal fine-grained details that might not be intuitively obvious.

---

No prior research has counted citations to articles separately by originating section (i.e., the section of the citing article that contains the in-text citation) but some have reported the distribution of in-text citations between sections. A relatively large-scale study of 433 empirical articles from 39 subjects found the most common structure to be five parts: Introduction; Literature Review; Methods; Results and Discussion; Conclusion (Lin & Evans, 2012). Nevertheless, investigations of article sections containing citations have usually focused on the IMRaD (Introduction, Methods, Results, and Discussion) four-section structure, as recommended in many medical journals (ICMJE, 2004), although a wide variety of structures are used in practice (e.g., Kanoksilapatham, 2005; Maswana, Kanamaru, & Tajino, 2015; Tessuto, 2015; Teufel, 1999). As reviewed below, the Introduction and Discussion may contain the most citations (Bertin, Atanassova, Gingras, & Larivière, 2016).

The current paper investigates, but does not fully evaluate, counting citations by section for science. A full evaluation would require human interpretation to assess whether section-based citation counts give more useful information than raw citation counts but this is difficult at the scale of all science because it would require experts from all fields passing judgements on articles within their broad areas that they may not have read before. Instead, this paper identifies necessary preliminary information, by assessing the extent to which section-based citation counts give substantial new and internally reliable information. The following research questions address gaps in prior research, by giving finer grained information than before in terms of more fields (22) and more sections (6) (RQ1,4) and by addressing different issues (RQ2,3). The six sections used follow the finding that the six-section structure is equally common as IMRaD and the six sections in various combinations account for an estimated 80% of empirical articles, and all may be present in most empirical articles, as discussed below (Lin & Evans, 2012).

1. **Which of the six standard article sections (Introduction, Background, Methods, Discussion, Results and Conclusion) generate substantial numbers of citations in each broad field?** This covers more fields and sections than prior research (seven journals and IMRaD: Bertin, Atanassova, Gingras, & Larivière, 2016). It is an important question because these sections are relevant for section-based citation counting in each broad field.

2. **Which other standard sections generate the most citations for articles that are cited in each of the six standard sections, in each broad field?** Since there are many different article structures and articles can be cited in multiple different sections, it is important to identify, within each broad field, which pairs of sections overlap in the articles that they reference. For example, if Introduction citations tend to also be cited within Background sections within a given field then there would be little point in counting these sections separately.

3. **Does counting citations from one section produce different results (in terms of article rankings) from total citation counts, for each section and broad field?** If section-based citation count rankings differ little from (total) citation count rankings in a field, then there would be little point in section-based citation counting in that field.

4. **Are there highly cited articles that attract most of their citations from only one of the six sections?** If some articles attract most of their citations from one section then counting citations from that section might give information that is more specific than total citation counting. This is addressed only for highly cited articles as a practical step because articles with low citation counts could attract all citations from a single section by chance.

## 2 Background

Since the goal of this paper has apparently not been addressed before, this section covers research that has reported the number of citations per section or that has investigated different types of citation.

### 2.1 *Article structure and rhetorical function*

Academic articles need to be structured to convey their message persuasively; learning how to structure writing is therefore an important task for researchers. Many previous studies have examined how this is achieved in various fields. Within the social sciences, for example, articles seem to set the context of a study, describe the niche filled by the new paper and then describe their contribution. Within each of these three tasks, there are multiple subtasks that may occur, such as identifying a gap in research as part of describing a niche (Swales 1990, 2004). There are differences between fields the content that needs to be included within an article, however, and the range of acceptable structures for this. For example, pure maths papers have a rather different structure, starting with an introduction, definitions and results, then proofs (often with examples), followed by concluding remarks that often do not conclude in the traditional sense (Kuteeva & McGrath, 2015).

As mentioned in the Introduction above, whilst the IMRaD structure is standard in some areas of science, it is nevertheless far from ubiquitous and is probably not the most common structure, even for empirical research. For example, six common section types were often found in 433 empirical articles from 39 subjects within engineering, science, social science and the humanities: Introduction (100%); Literature review (51%); Method (89%); Results (49%); Discussion (52%); Conclusion (74%). The most common structures found were: ILM[RD]C (21%); IM[RD]C (16%); IMRDC (12%); IMRD (12%); and ILMRDC (12%); ILMRD (7%); others (20%) (Lin & Evans, 2012).

Sections of text in a paper may be identified that perform a range of common roles. One small scale analysis found several "zones" to occur frequently in both cardiology and computational linguistics research: Background; Topic; Related work; Purpose/problem; Solution/method; Result; Conclusion/claim (Teufel, 1999: p. 108). Other similar classifications have been proposed, such as for biochemistry and chemistry: Background; Hypothesis; Goal; Motivation; Object; Method; Model; Experiment; Observation; Result; Conclusion (Liakata, Saha, Dobnik, Batchelor, & Rebholz-Schuhmann, 2012). There are also schemes that include the relationship of the authors to the information, such as whether they are the originators of the knowledge claims made in the article (Teufel, Siddharthan, & Batchelor, 2009).

Whilst important rhetorical tasks might be flagged by separate section headings, this is not always necessary (Swales 1990, 2004; Teufel, 1999; Teufel & Moens, 2002). Thus, section headings might be guides rather than definite statements of content.

### 2.2 *Sections containing citations*

Two studies have investigated citations by section for individual information science journals. The citations in 350 articles with 3-6 sections from the Journal of Informetrics have been investigated, finding that the first section tends to contain the most citations (Hu, Chen, & Liu, 2013). In the Journal of the American Society for Information Science and Technology (JASIST), literature review sections, when present, unsurprisingly contained the most citations, of the six standard sections (Ding, Liu, Guo, & Cronin, 2013).

On a much larger scale, for 47362 articles in seven PLOS journals organised into IMRaD, the introduction contained the most references (peaking just after the start) and the discussion contained the second most references (peaking in the middle). The four components contained similar numbers of references in each journal. The Results and Discussion tended to have slightly younger references than the Introduction, with the Methods containing the oldest references (Bertin, Atanassova, Gingras, & Larivière, 2016).

For all scientific fields and tens of thousands of journals, another study analysed six million publications from the PMC Open Access collection (journal articles or reviews, published from 1998 and gathered in October 2015) and the set of all Elsevier journals ('full-length article', 'short communication', or 'review article' accessible in Leiden University, from 1998 onwards, gathered in January 2017). It did not investigate individual sections but instead analysed the position of each reference within the body of an article. It found the distribution of references (number of citations per reference) within the text of Elsevier articles from the three broad fields of Biomedical and Health Sciences, Life and Earth Sciences, and Physical Sciences and Engineering had the same shape (peak value about 35% through the article; high values at the start and the end of the article), whereas the distribution of citations per reference within the text of articles from the remaining two broad fields had two different shapes: Mathematics and Computer Science (relatively flat distribution; peak at 60%) and Social Science and Humanities (peak at 65%) (Boyack, van Eck, Colavizza, & Waltman, 2018).

## 2.3 Citation contributions

A citation can perform different functions, including setting the background of a study, explaining and justifying methodological choices, and helping to incorporate the findings of new research into the knowledge of a field. There have been many studies of citation contributions but all have been based on a limited number of articles in few fields (usually 1-3). This is an important limitation because citation practices vary between disciplines (Hyland, 1999).

Citation analysis rests on the assumption that citations tend to acknowledge relevant prior contributions to knowledge (Merton, 1973), but a study of cardiology and diabetes citations estimated that only 1% were essential to the citing paper and that most citations made a limited contribution (Hanney, Grant, Jones, & Buxton, 2005). A citation can be perfunctory because its role seems to be to acknowledge that prior similar work has been conducted (Cano, 1989). Authors may also choose to include citations for indirect reasons, such as for self-promotion or because their field expects a literature review (Vinkler, 1987), and may be influenced by ostensibly irrelevant factors, such as language (Lillis, Hewings, Vladimirou, & Curry, 2010). A study of structural engineering citations found that a quarter were important for understanding the new work, with a fifth each relating to a theory or concept (19%) or methods (23%) used in the citing article (Cano, 1989). Another small study similarly differentiated between "knowledge" and "experimental protocol" citations for biomedical research (Yu, Agarwal, & Frid, 2009). A follow up study used a categorisation scheme that mirrors to some extent the text purposes discussed above, "Background/Perfunctory, Contemporary, Contrast/Conflict, Evaluation, Explanation of results, Material/Methods, Modality, and Similarity/Consistency" (Agarwal, Choubey, & Yu, 2010).

From the perspective of the contribution to the citing article, three relevant types of citation are: perfunctory, theory used and methods used. Whilst all these types of citation could occur anywhere in an article, the logical place for methods citations is in the Methods or a related section, and theory citations would presumably be likely to occur in Introduction or Background sections in the early part of an article. Thus, citations in different parts of an article are likely to play different roles (Suppe, 1998).

Previous research has confirmed that the contribution of a citation is influenced by the section containing it, although mostly based on small samples. The Introduction has been found to contain the most highly cited articles in an early study (Voos & Dagaev, 1976) and both the Introduction and Literature Review sections in another (Ding, Liu, Guo, & Cronin, 2013). Methods-related citations have been shown to dominate many highly cited papers lists (Small, 2018) and if these citations tend to originate from Methods sections then the Methods section may also contain the most cited articles in some fields. An investigation of citations in the Methods sections of 32 language education journal articles found that they were often used to justify the materials and methods used, for example (Miin-Hwa Lim, 2014). Two other analyses judged that citations from the Introduction section were less important than those in the Methods, Results and Discussion (Maričić, Spaventi, Pavičić, & Pifat-Mrzljak, 1998; Tang & Safer, 2008). For example, introductory citations may not be necessary to understand or justify an article, or there may be selected in preference to other citations that may have performed the same function. Another study analysed Literature Review sections separately from Background sections and argued that citations in both had little importance for library and information science research (Zhao, Cappello & Johnston, 2017). Combining these two sets of studies creates the suggestion that the most highly cited papers do not necessarily make the most important contributions to future research.

Another marker of citation contribution is whether it occurs separately or as part of a list of citations within the text. A citation within a list may be regarded as not making a substantial contribution or being redundant (Bonzi, 1982; Moravcsik & Murugesan, 1975).

A reference may be cited multiple times within an article, with a survey of three fields finding that more frequently mentioned articles were judged to make a greater contribution (Tang & Safer, 2008; Zhu, Turney, Lemire, & Vellino, 2015). Moreover, if an article is cited in multiple different sections then this may indicate a more substantial contribution (Herlach, 1978). Articles cited only once in a text are more likely to be highly cited (Boyack, van Eck, J., Colavizza, & Waltman, 2018), however, confirming that highly cited papers may contribute relatively little each time they are cited. There are substantial field differences in the likelihood that an article is cited multiple times within a text and this is changing over time (Boyack, van Eck, Colavizza, & Waltman, 2018) but it is not clear why.

## 3   Methods

The PubMed open access XML article collection was used as the dataset, as downloaded in October 2018. It included 2,177,956 documents from 8,525 journals. This is apparently the largest free full-text collection of structured scientific articles. It contains documents deposited by open access journals as well as individual open access papers within hybrid journals. It has a biomedical focus, but contains substantial numbers of articles in most areas of academia, except for the arts and humanities.

The downloaded articles are in XML format, obeying National Information Standards Organization (NISO) Journal Archiving and interchange Tag Sets (JATS) standards. Articles must report a type (e.g., research-article, review-article, other) in a metadata tag. They may also report section names, either in the "sec-type" attribute of the "sec" tag, or in a title tag following the section start tag. These were used to determine section names. Article XML files were processed to extract meta-data and documents that were not declared as type research-article were discarded to avoid contaminating the results with non-standard document headings, giving 1,584,674 research articles from 8222 journals. Articles were split into sections using the section tags, with the section type label within the section XML tag (when present) or the initial title tag to name the section. Sections may be nested within articles (e.g., 2 Methods; 2.1 Data; 2.2 Analysis) and only the outer section was used in each case. Common similar section titles were standardised to avoid minor variations influencing the results. In particular, the Methods section included sections labelled Material[s] and Methods, Statistical Analysis, or Data, and the Background section also included sections labelled Literature Review. Initial section numbers were removed before checking title names. The following names were used to identify each section. These include some tag names that are not displayed in the article. The list was obtained by examining the most common section names and selecting unambiguous and relatively general titles.

- Introduction: intro; introduction.
- Background; background; literature review; related literature.
- Methods: materials|methods; methods; materialsandmethods; materials; statistical analysis; data analysis; statistical analyses; statistics; study design; study population; data collection; procedure; statistical methods; measures; patients and methods; data; experimental design; research design and methods; data extraction; sample collection; experimental procedures; methods/design.
- Results: results.
- Discussion: discussion; results and discussion; limitations; strengths and limitations; study limitations.
- Conclusion: conclusion; conclusions; summary; concluding remarks; summary and conclusions; summary and conclusion; conclusions and outlook; conclusions and perspectives; conclusions and recommendations; conclusion and perspectives; conclusion and outlook; conclusions and future work; conclusion and future work.

Reference lists were extracted from the XML sections at the end. Each reference is given a tag by which it can be referred to by in-text citations, allowing the positions of each reference in the text to be detected. Most (87.6%; 53,619,183 out of 61,229,921) of the cited references were recorded as being type "journal", although no finer grained reference scheme was used. Only references with DOIs were retained (32.3%; or 19,748,147 out of 61,229,921) to avoid errors due to false matches. Few references included a PubMed ID within the cited references XML (0.3%; 157,160 out of 61,229,921) but these were not used because for some of the analyses it was important to track the same article being cited in different fields, so a single unique identifier was needed. It seems likely that social science, arts and humanities fields would be less likely than natural sciences to include DOIs in references. Moreover, this inclusion is partly a journal policy. Because of these factors, the use of DOIs means that absolute numbers of references per article cannot be reliably compared between fields (although this is not relevant here). This should allow near perfect matching so that patterns in the results could not be caused by incorrect matches. Articles are sometimes implicitly cited in lists of references (e.g., "[3]-[6]") and rules were developed to detect the standard ways in which this occurred.

For each article, a list of referenced DOIs was generated for each section and these lists were then used to create a citation count for each DOI, recording separately the number of citations from each section. When articles were mentioned in multiple sections, they were given fractional citation counts corresponding to the number of mentions in the section and overall. For example, an article mentioned twice in the Introduction and once in the Discussion would have an Introduction citation count of 2/3 and a Discussion citation count of 1/3. Citations were ignored if they were in sections that were not named, or given a name that was not resolved into one of the six investigated, or if they were not within a section.

For a disciplinary analysis, the DOIs needed to be separated by broad field. The public Science-Metrix journal classification list (22 categories) was used for this (Archambault, Beauchesne, & Caruso, 2011), with extensions for large reputable scholarly journals that had been omitted. The Science-Metrix scheme puts journals into separate categories, in contrast to several other major journal classification schemes, which is an advantage and seems to be more accurate (Klavans & Boyack, 2017). After removing articles in journals that were not in the Science-Metrix scheme, 799,055 articles had at least one citation in the six sections combined. Note that the cited articles can be of any type (with a DOI and in a journal) even though the citing articles are all declared to be research articles.

The 22 fields vary greatly in the number of articles included, from 10 (Visual & Performing Arts) to 341,544 (Clinical Medicine), as shown in the Appendix. The Visual & Performing Arts category was excluded from the main analyses due to its small sample size. Data for fields with lower numbers of articles is less reliable. This affects most of the results for Visual & Performing Arts and the precision of some results for other smaller fields. These issues are commented on, when relevant in the Results.

For RQ1, for each field and section, the average number of citations per article was calculated using the geometric mean (and corresponding 95% confidence intervals) with a one offset because of the presence of zeros in the data (Thelwall & Fairclough, 2015). This is more appropriate than the arithmetic mean due to the skewed nature of citation counts. This is a zero-truncated calculation because it excludes all uncited articles. The information it gives is therefore about cited articles rather than articles in general. The cited articles are from different years and since citation counts vary by year and Methods citations tend to be older (Bertin, Atanassova, Gingras, & Larivière, 2016), this exaggerates the importance of methods citations.

For RQ2, the geometric mean citation calculation above was repeated for subsets of the data, such as the subset of articles with at least one citation in the Introduction. The average citation counts for all six sections were then compared graphically for articles with at least one citation in each section.

For RQ3, Spearman rank correlations were calculated for citation counts overall and each section, again separately for each field. Because the data is zero truncated, the correlations are likely to be weaker than if the dataset included many articles that had no citations from any source. Because the cited articles are from different years, this gives a positive skewing on the correlations because older articles are likely to be more cited overall. Thus, the correlations were calculated only for cited articles published in one year, 2012. This gives a citation window of at least five years for each article, which is a reasonable period for articles to have been cited. A relatively large citation window was useful to ensure that most articles would have reached close to their maximum citation counts.

For RQ4, the two articles with the highest share of citations from each of the six sections were investigated to assess whether their unusual position reflected section-specific types of impact. This is a small-scale investigation to gain insights into section-specific impacts. A larger scale systematic assessment would be needed to give robust evidence. Only articles with at least 100 citations were included to reduce the likelihood that the large share of citations from one section did not have a systematic cause.

# 4   Results

Data underlying Figures 1 to 8 can be found in the online supplement.

## 4.1   RQ1: Sections generating the most citations

The average number of citations per section was generally lowest in the Methods, Results and Conclusions sections (Figure 1). In some broad fields, such as Earth & Environmental

Sciences, there were few Background section citations because of the tendency to avoid this section name (or standard name variants) but to put background information in the Introduction. Fields in which there were most citations in the Discussion may have shared background citations between the Introduction and Background sections that were then returned to in the Discussion section combined, or may have introduced background material in the Discussion for the first time to contextualise the results, lacking an earlier substantial discussion of background material. The following general patterns can be observed, in terms of the sections hosting the most citations. Only the second pattern (Introduction and Discussion) encompasses a set of related fields and so there is not a simple explanation of the results in terms of very broad field publishing strategies.

- **Introduction**: The Introduction hosted the vast majority of citations in eleven different areas (Built Environment & Design; Communication & Textual Studies; Earth & Environmental Sciences; Economics & Business; General Arts, Humanities & Social Sciences; General Science & Technology; Historical Studies; Mathematics & Statistics; Physics & Astronomy; Social Sciences; Visual & Performing Arts [unreliable]).
- **Introduction and Discussion**: The Introduction and Discussion both contain substantially more citations than the other sections in three life science related areas (Agriculture, Fisheries & Forestry; Biology; Biomedical Research). The Introduction contained more citations than the Discussion for Psychology & Cognitive Sciences and less citations for Clinical Medicine, but otherwise these two fields also fit the same pattern.
- **Introduction and Background**: The Introduction and Background both contain substantially more citations than the other sections in Chemistry.
- **Three or more sections**: The remaining fields had substantial numbers of citations from three (Philosophy & Theology; Public Health & Health Services), four (Engineering; Information & Communication Technologies) or five (Enabling & Strategic Technologies) different sections, although never from the Conclusions. The mix of sections and their relative weights is not the same for any two fields.

The citations were also analysed from the perspective of the field that *received* the most citations (Figure 2). This gives a radically different perspective. The largest change is that the Introduction is not the dominant section. This is because disproportionately many citations from the Introduction do not target documents with DOIs. The missing citations are likely to include a large share from classic older papers that lack DOIs. This is because, as discussed above, the Introduction can play a context setting role. The second change is that there are no obvious patterns in the graphs. Although there are some similar graph shapes, since there are many different shapes and the most similar are from different areas of scholarship, it does not seem reasonable to claim the existence of any patterns. For example, the dissimilar Clinical Medicine and Economics & Business have similar numbers of citations from the Introduction, Background, Methods and Discussion but few from the Results and Conclusions. Moreover, four contrasting fields have similar method-dominated graphs (Biology; Enabling & Strategic Technologies; Mathematics & Statistics; Social Sciences).

The Results and Conclusion were not the main sources of citations in any target field. The surprisingly high values for the Methods section is presumably be due to fields with little representation in PMC (hence few within field disciplinary citations) but providing methods for other fields. The following lists the section generating the most citations for each field, emphasising the variety in the data. The inclusion of the Background and the exclusion of the Results is perhaps most significant in terms of not following the recognised IMRaD structure in some fields.

- **Introduction**: Built Environment & Design; Chemistry; Communication & Textual Studies; Earth & Environmental Sciences; Engineering; Historical Studies; Psychology & Cognitive Sciences.
- **Background**: Clinical Medicine; General Arts, Humanities & Social Sciences; General Science & Technology.
- **Methods**: Biology; Biomedical Research; Enabling & Strategic Technologies; Information & Communication Technologies; Mathematics & Statistics; Physics & Astronomy; Public Health & Health Services; Social Sciences.
- **Discussion**: Agriculture, Fisheries & Forestry; Economics & Business; Philosophy & Theology.

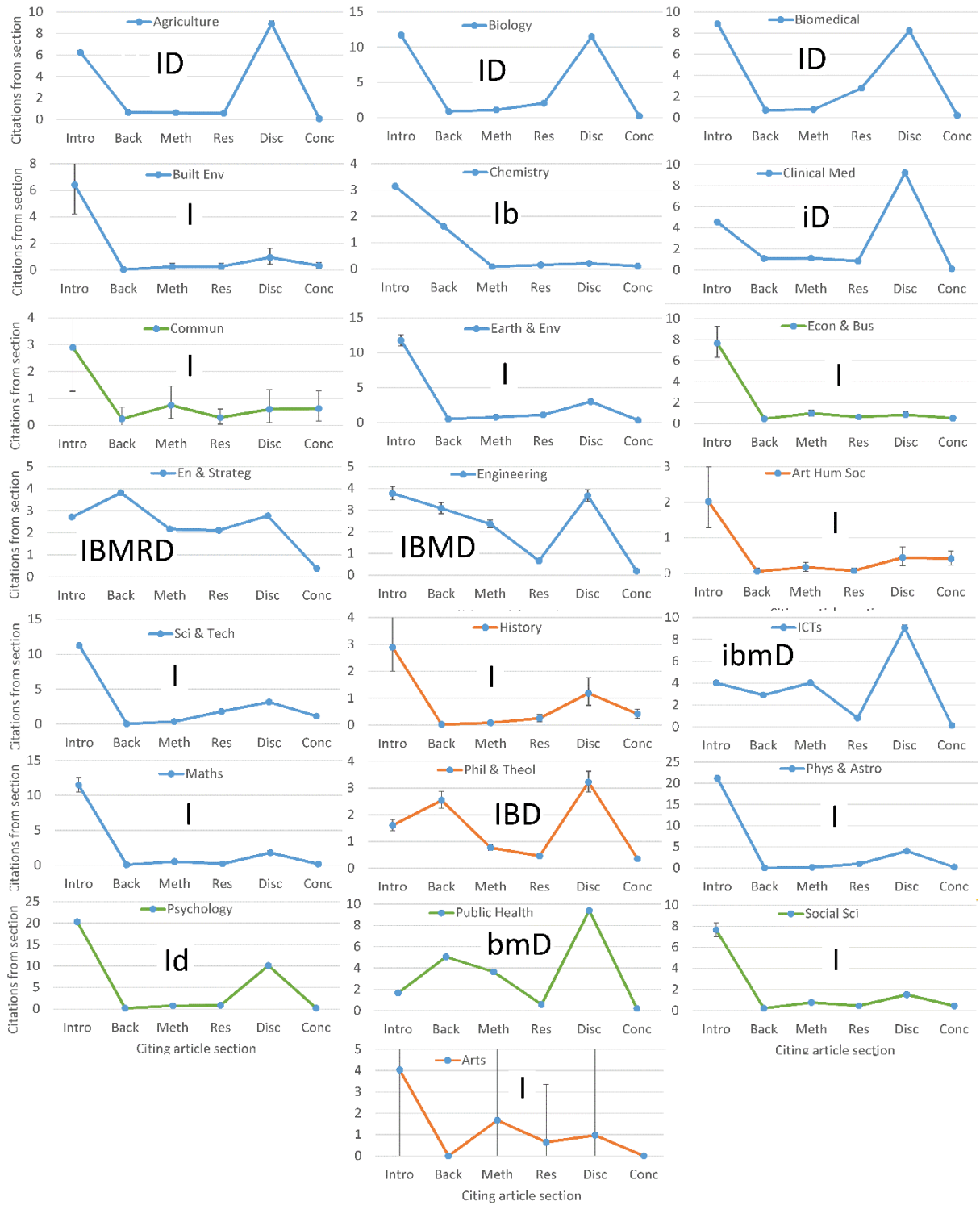

Figure 1. Share of article citations from six common sections of PMC research articles, by **source** article Science-Metrix broad class. Qualification: source article has DOI. Sample sizes are in the Appendix, Table A1 and Table A2. Arts and Humanities subjects are orange and social sciences are green (History is partly both). Letters indicate the main sections hosting citations, with lower case indicating fewer citations than the main section.

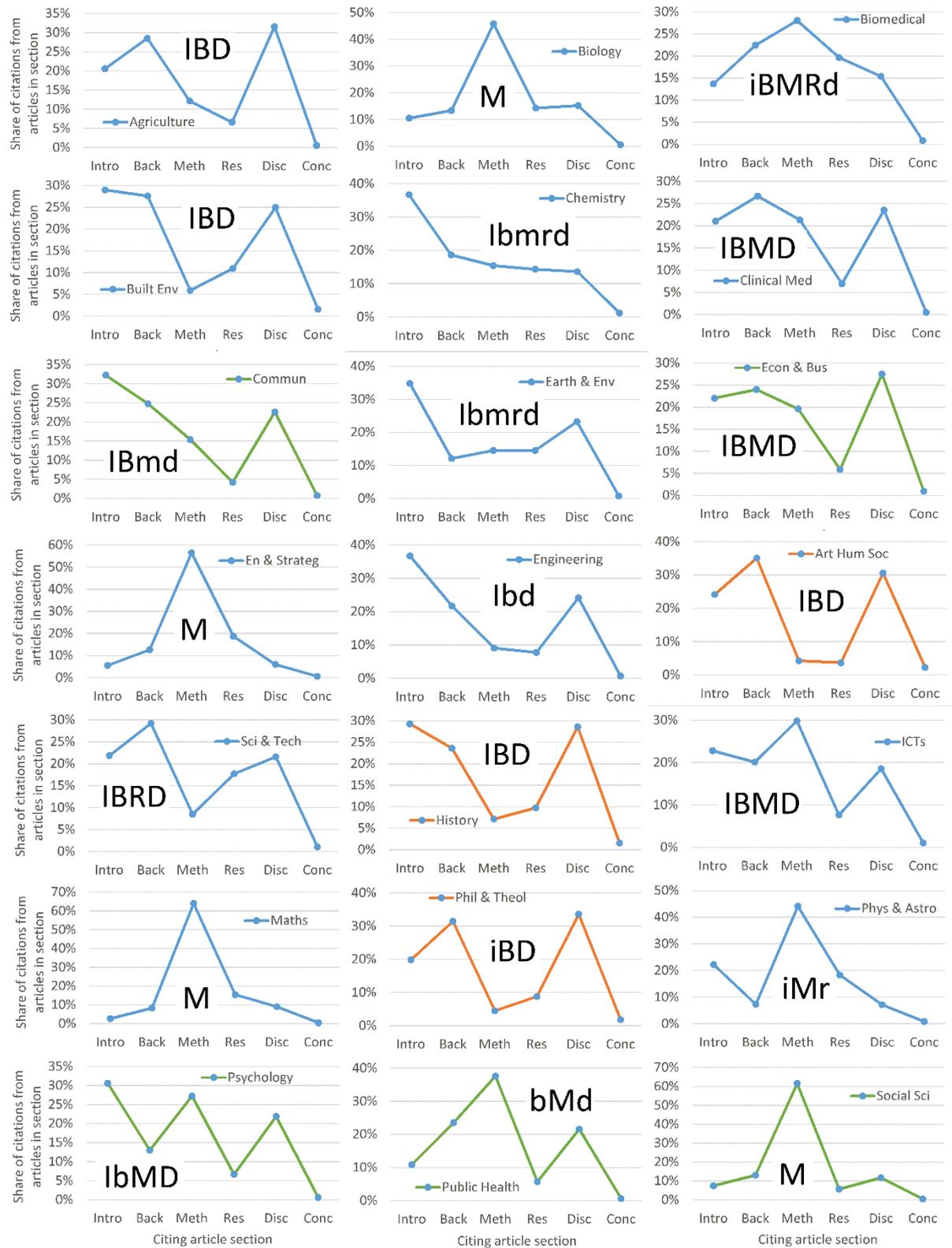

Figure 2. Share of article citations from six common sections of PMC research articles, by **target** article Science-Metrix broad class. Qualification: target article has DOI. Sample sizes are in the Appendix, Table A1 and Table A2.

## 4.2 RQ2: Sections generating the most citations for articles cited in each section

For articles cited in each section, the other sections also citing the same article were investigated. When there is almost no overlap (i.e., articles cited by the section are rarely cited by other sections) this indicates section-specific impact for the article. When there is an overlap, then either the article has multiple different types of impacts or there is an overlap in the purpose of different citing sections.

**Introduction** (Figure 3): In eight fields, articles cited in the introduction are rarely cited in other sections, so Introduction citations play a relatively unique role. Articles that are cited in Introductions are most likely to be also cited in the Discussion, with the Background being the second most likely section. The overlap between the Introduction and Background is presumably due to literature reviews being included in introductions and backgrounds for different articles in some areas.

**Background** (Figure 4): Background section citations play a relatively unique role in four fields but in most they have overlaps with Introduction and/or Discussion citations. Mathematics & Statistics is unique in having an overlap with citations from Methods sections.

**Methods** (Figure 5): Methods-cited articles are relatively unique in eight fields. In others, they can also be cited in any section except the Conclusions. Citations in other sections could be due to methods discussions elsewhere in empirical articles, articles with a focus on methods, or theoretical contributions to methods sections.

**Results** (Figure 6): Results-cited articles are not unique in any field. They are closest to unique in three fields with a degree of overlap in subject matter: Chemistry; Earth & Environmental Sciences; Engineering. Results-cited articles are also cited in the Introduction, Background and/or Discussion in many fields.

**Discussion** (Figure 7): Discussion-cited articles are relatively unique in three unrelated fields: Agriculture, Fisheries & Forestry; Communication & Textual Studies; Historical Studies. In most other cases there is an overlap with the Introduction and/or Background.

**Conclusions** (Figure 8): Conclusion-cited articles are not unique in any field and in most articles cited in the Conclusions are usually more cited by another section. Thus, it seems likely that being cited in the Conclusions does not have a special role in any large area of scholarship. Presumably this is because Conclusions tend not to introduce new material except as deduced from the analysis earlier in a paper.

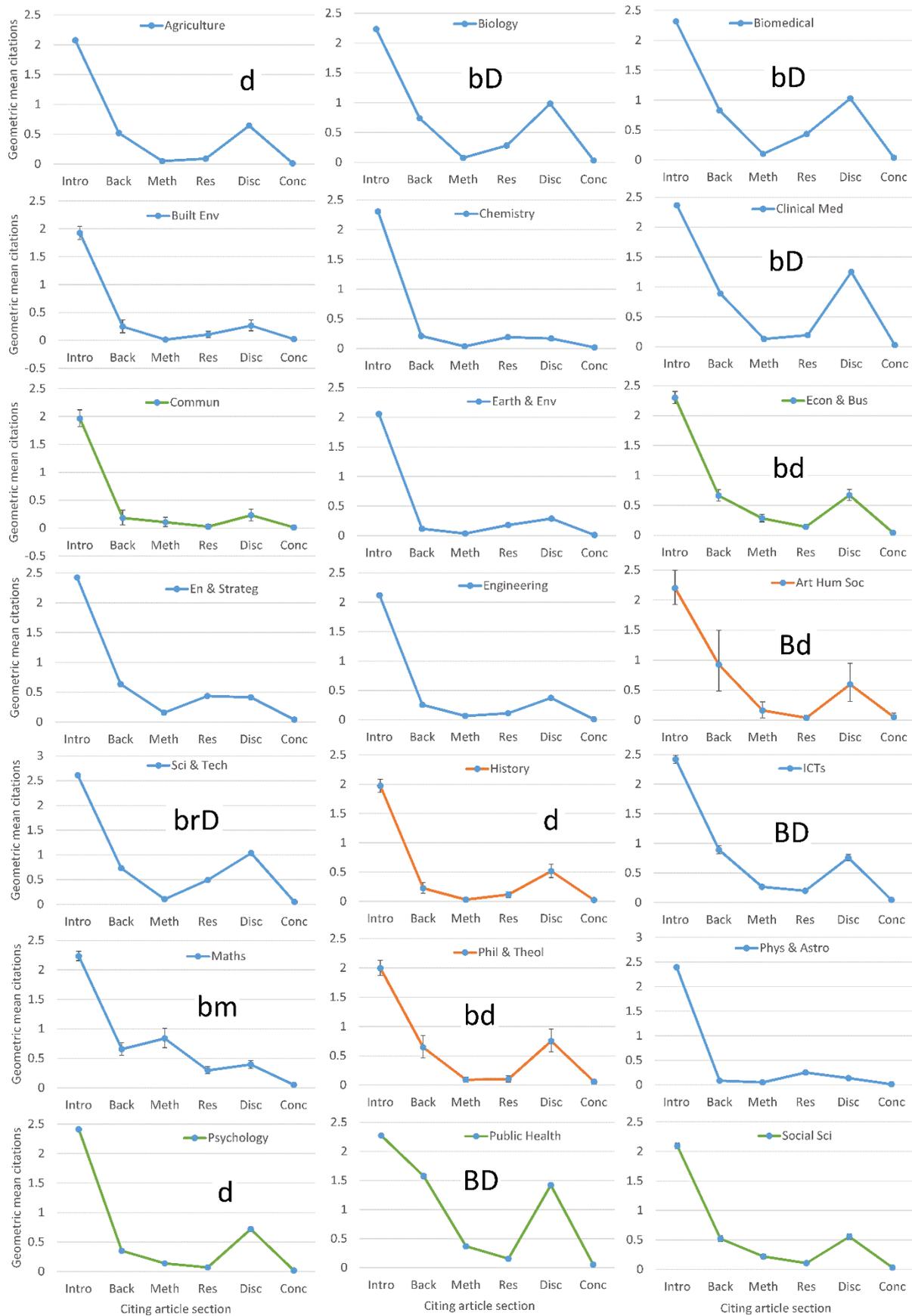

Figure 3. Average (geometric mean) number of citations **per article** from six common sections of PMC research articles, by target article Science-Metrix broad class. Only for articles cited at least once in an **Introduction**. Sample sizes are in the Appendix, Table A1 and Table A2.

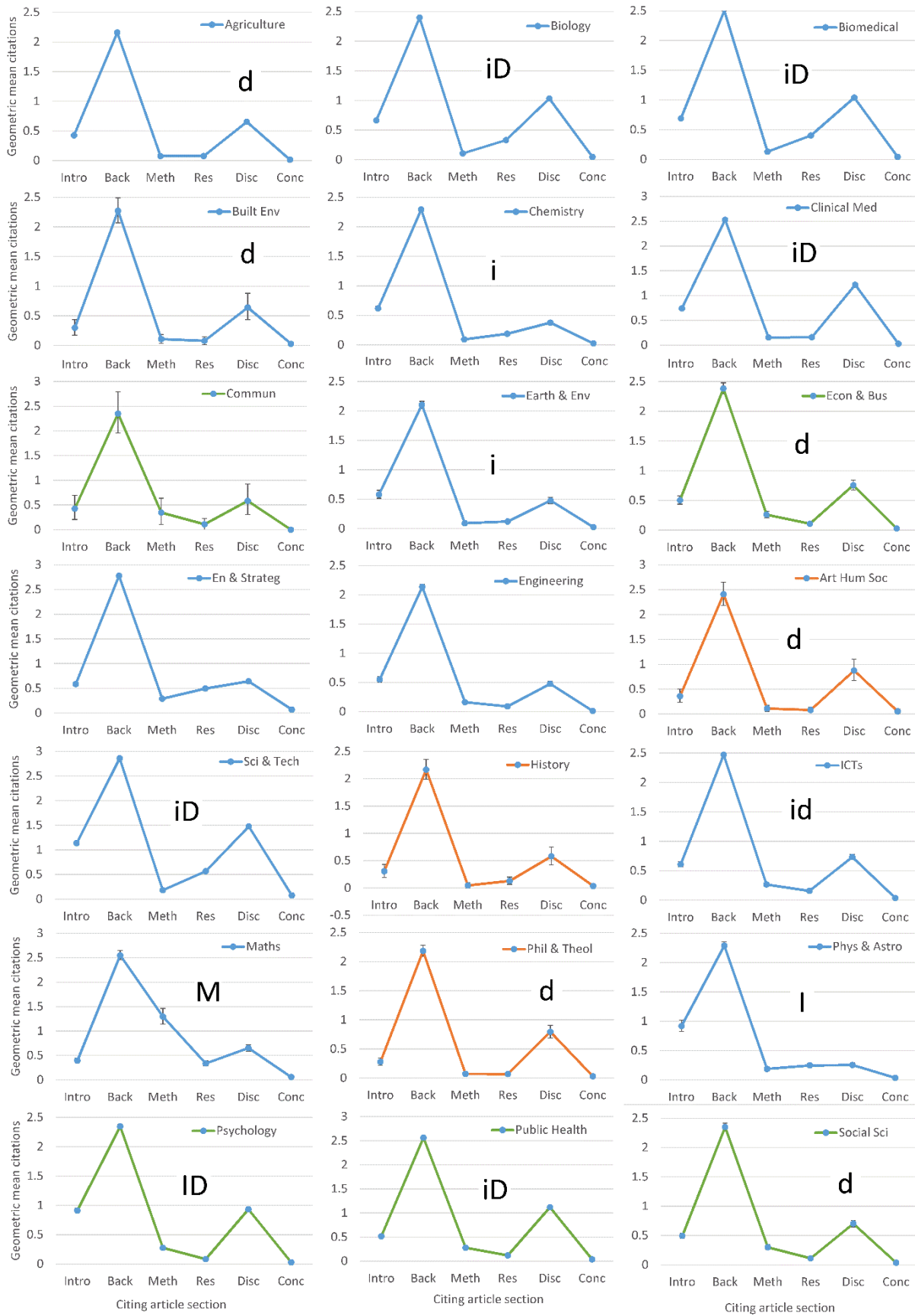

Figure 4. Average (geometric mean) number of citations **per article** from six common sections of PMC research articles, by target article Science-Metrix broad class. Only for articles cited at least once in a **Background**. Sample sizes are in the Appendix, Table A1.

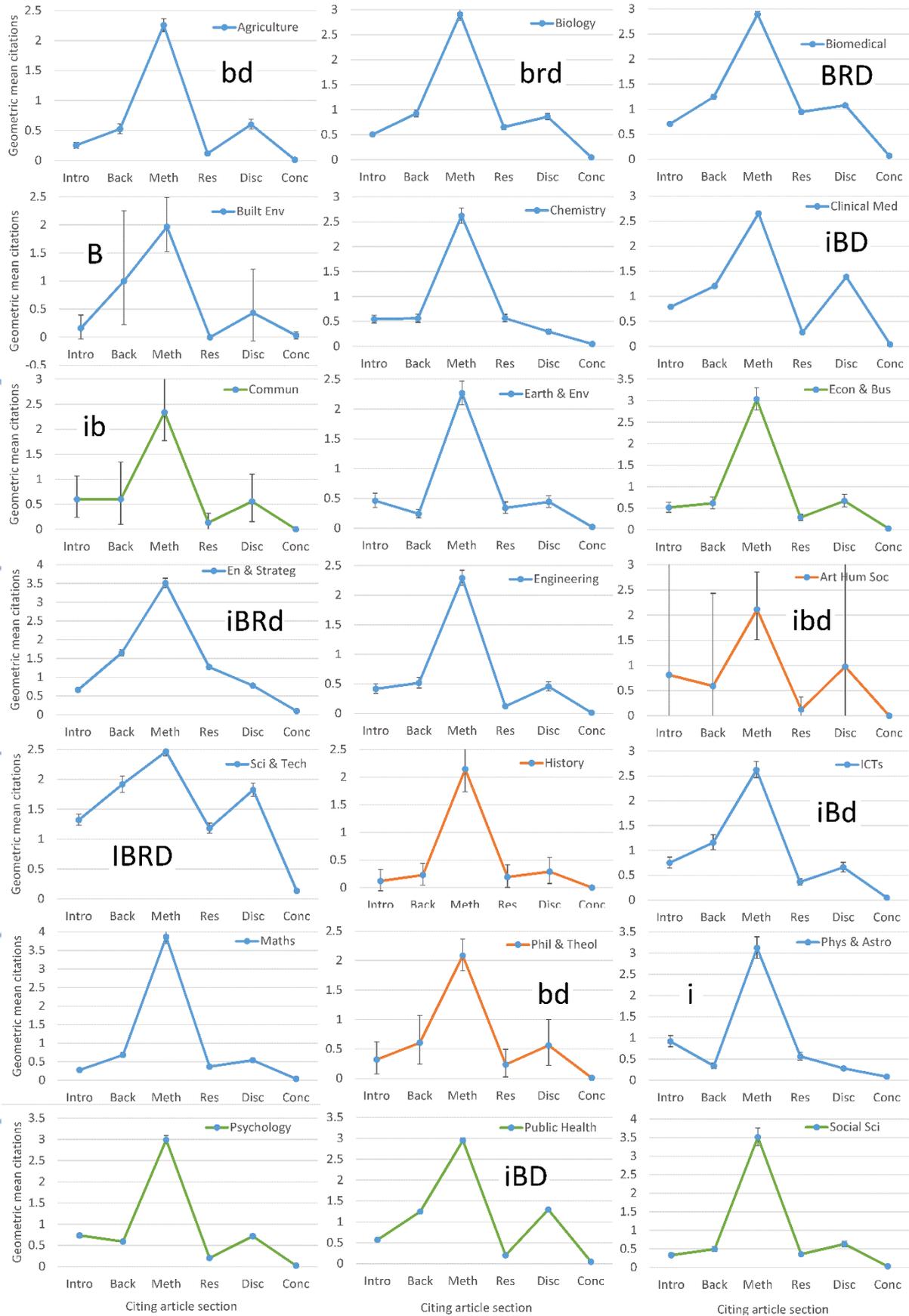

Figure 5. Average (geometric mean) number of citations **per article** from six common sections of PMC research articles, by target article Science-Metrix broad class. Only for articles cited at least once in a **Methods** section. Sample sizes are in the Appendix, Table A1.

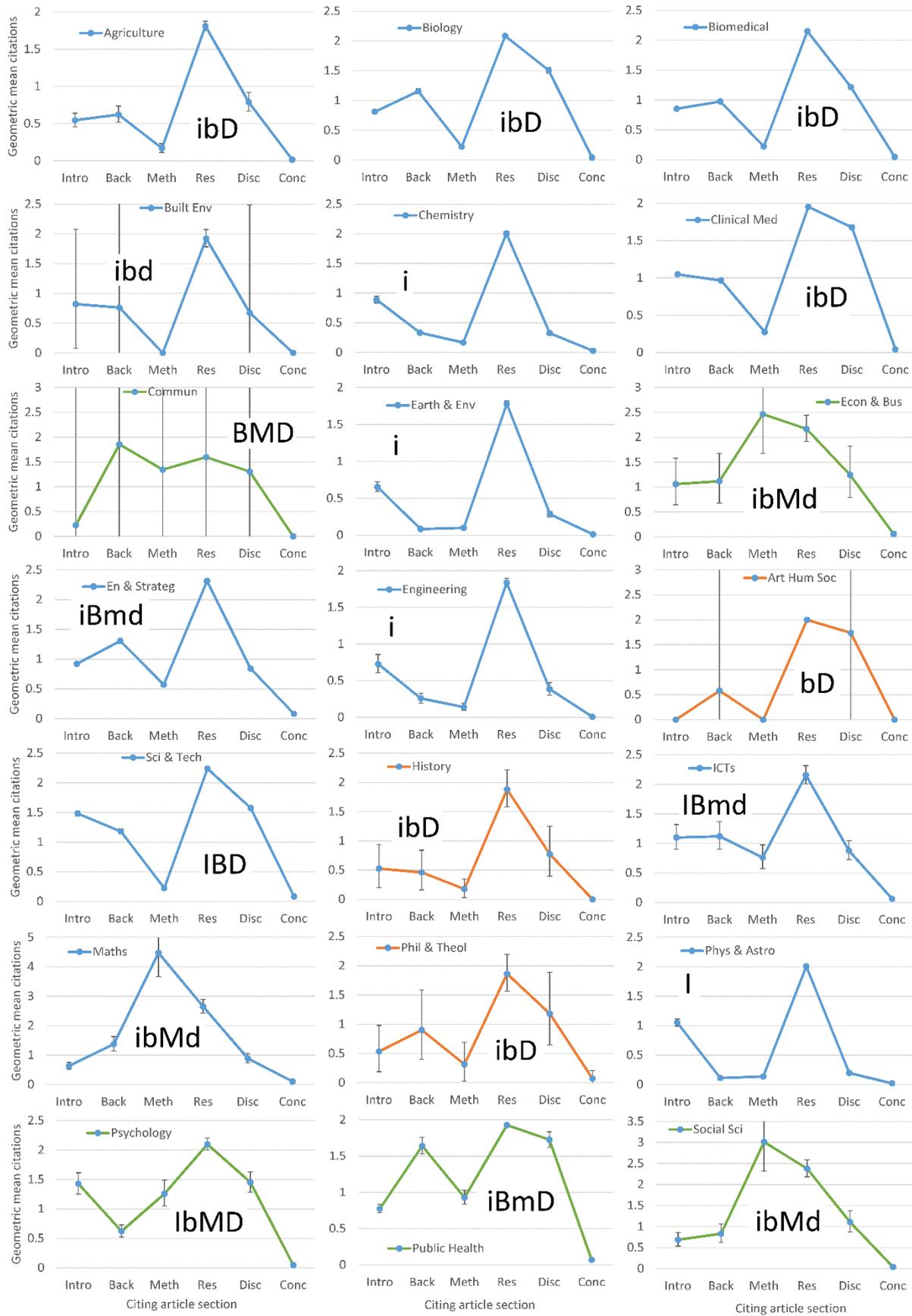

Figure 6. Average (geometric mean) number of citations **per article** from six common sections of PMC research articles, by target article Science-Metrix broad class. Only for articles cited at least once in a **Results** section. Sample sizes are in the Appendix, Table A1.

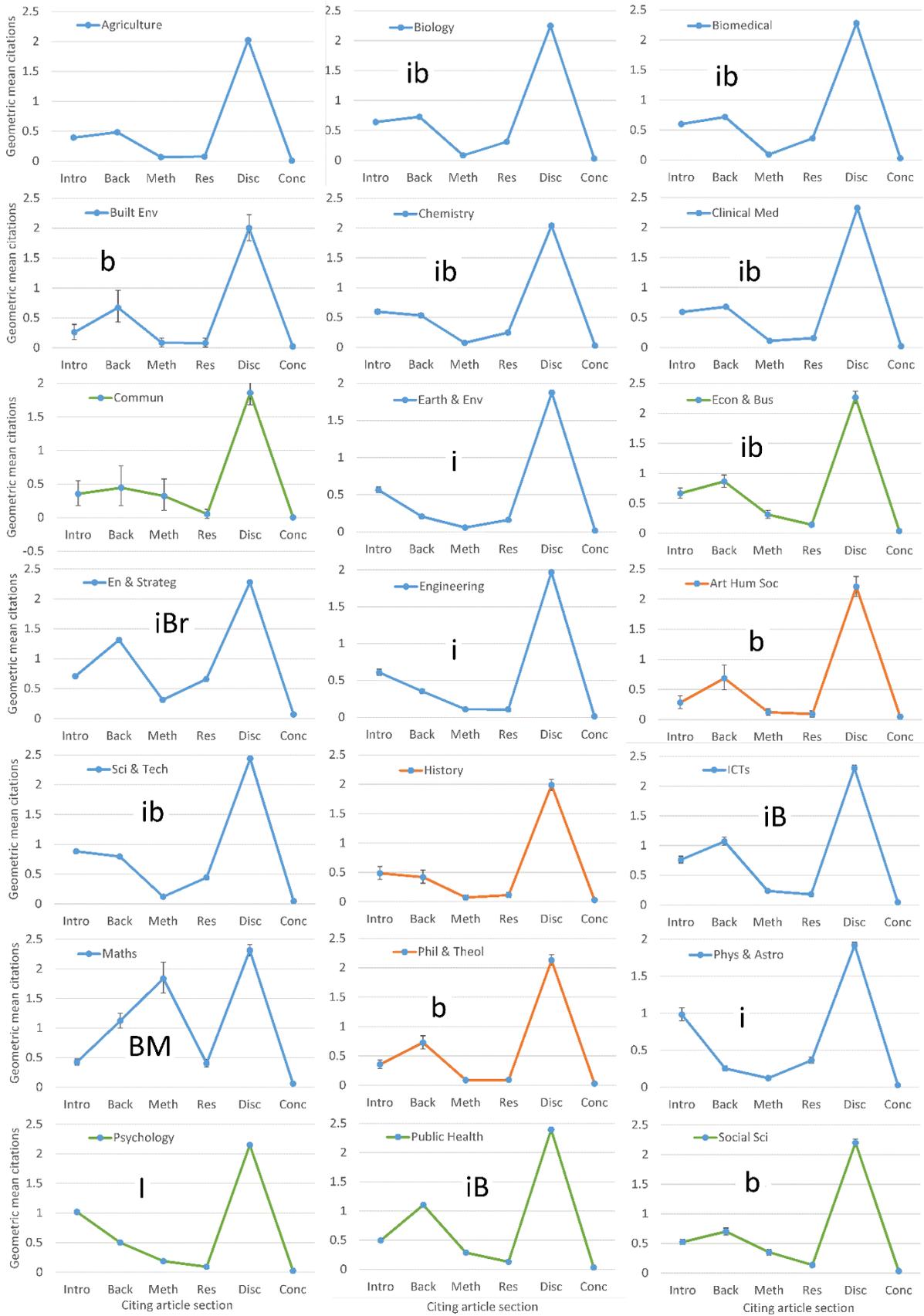

Figure 7. Average (geometric mean) number of citations **per article** from six common sections of PMC research articles, by target article Science-Metrix broad class. Only for articles cited at least once in a **Discussion**. Sample sizes are in the Appendix, Table A1.

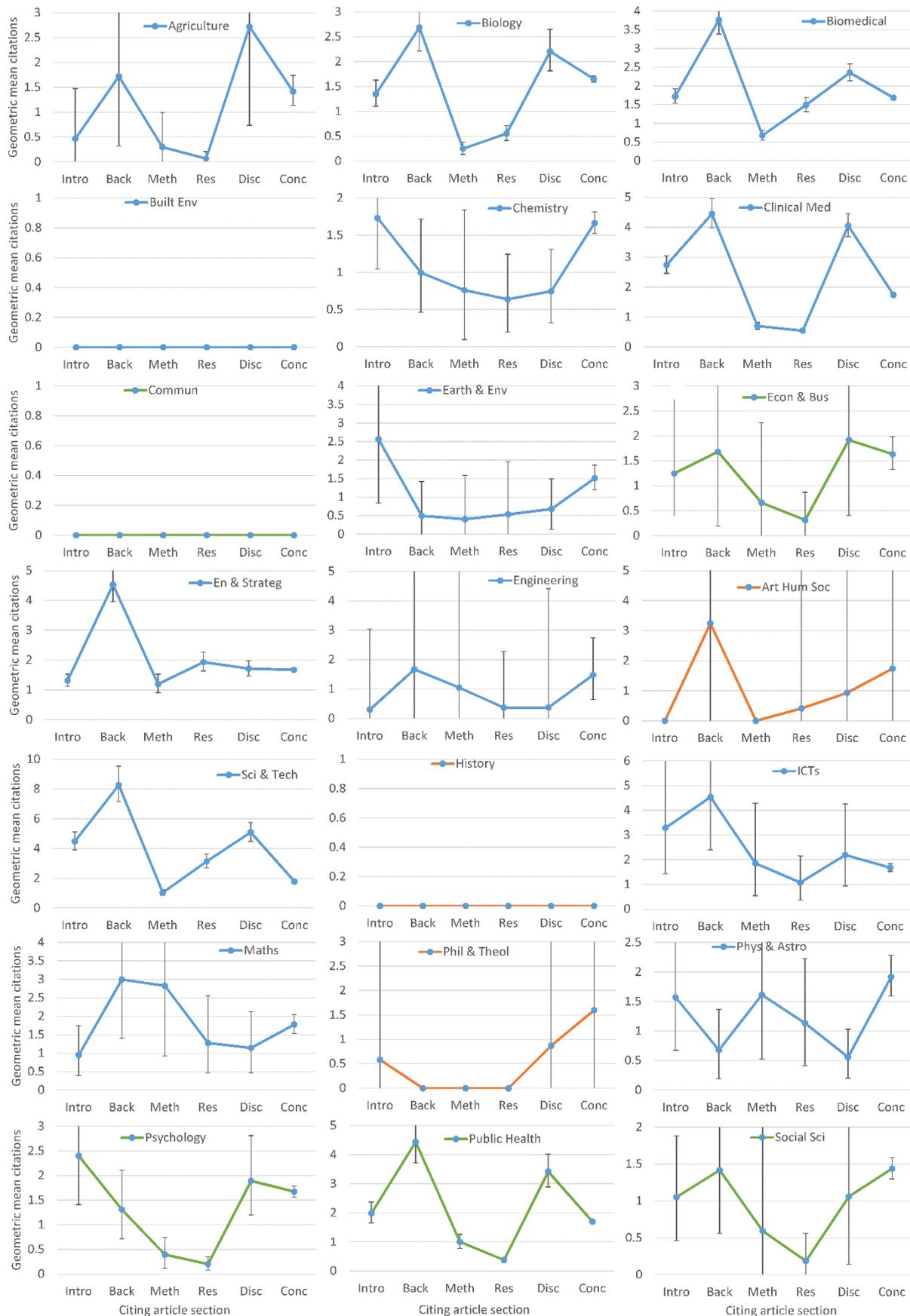

Figure 8. Average (geometric mean) number of citations **per article** from six common sections of PMC research articles, by target article Science-Metrix broad class. Only for articles cited at least once in a **Conclusion**. Sample sizes are in the Appendix, Table A1.

## 4.3  RQ3: Ranking citation counts by each section

The rank order of articles changes dramatically if citations are counted from only one of the six sections rather than from all sections (Table 1; see also Appendix Table A1 for all articles). The smallest rank order change (correlation: 0.44) would occur if citations were counted only from the Discussion. The correlation between citation counts from different pairs of sections are mostly close to zero. The correlations would presumably be higher if the data set was not zero truncated (excluding articles not cited in any of the five sections) due to the inclusion of many matched zeros (zero overall and zero in each section).

The negative overall correlation between citations from the Introduction and citations from the Background section (Table 1) and a negative correlation in nearly all fields (18 out of 21: Table 2; see also Appendix Table A3 for all articles and Appendix Table A4) is probably due to field differences in whether a separate Background section is used or whether background material is included in the Introduction. References that are frequently cited by articles of the first type tend not to be frequently cited by articles of the second type. References cited in the Introduction also tend not to appear in the Discussion (Table 1, Table 2). This contrasts with Figure 3 and to some extent Figure 7, which have relatively high values for both Introduction and Discussion, but the figures do not take into account the relatively high number of articles cited by both sections (Figure 1). For example, even if many articles that are cited in Discussions are also cited in Introductions, more that are cited in Discussions are not cited in Introductions and vice versa.

Two areas have low median positive correlations but nevertheless have a positive correlation in nearly all fields (18 out of 21): Methods and Results; and Results and Conclusions. The former case suggests that Results references often relate to methods, whereas the latter case suggests that results references are often returned to in Conclusions. Alternatively, different articles with similar topics may include methods details in the Results in the former case.

Table 1. Median correlation between citation counts calculated only from the named section (n=21 fields: no articles for Visual & Performing Arts; median sample size 552) and for cited articles published in 2012.

| Section | Back | Meth | Res | Disc | Conc | **All** |
|---|---|---|---|---|---|---|
| Introduction | -0.15 | -0.10 | -0.04 | -0.09 | 0.03 | 0.37 |
| Background | 1.00 | 0.02 | -0.03 | 0.00 | 0.00 | 0.31 |
| Methods | | 1.00 | 0.08 | -0.02 | 0.04 | 0.20 |
| Results | | | 1.00 | -0.01 | 0.04 | 0.18 |
| Discussion | | | | 1.00 | 0.02 | 0.44 |
| Conclusions | | | | | 1.00 | 0.09 |

Table 2. Number of fields with a positive correlation between citation counts calculated only from the named section (n=21 fields) and for cited articles published in 2012. High and low percentages are bold.

| Section | Back | Meth | Res | Disc | Conc | All |
|---|---|---|---|---|---|---|
| Introduction | **14%** | **19%** | 38% | 14% | 71% | **90%** |
| Background | 100% | 62% | 24% | 52% | 62% | **100%** |
| Methods | | 100% | **86%** | 38% | 76% | **90%** |
| Results | | | 100% | 48% | **86%** | **90%** |
| Discussion | | | | 100% | 71% | **100%** |
| Conclusions | | | | | 100% | **86%** |

### 4.4 RQ4: Highly cited articles attracting most citations from one section

Some highly cited articles attracted most of their citations from one section type (Figure 9). This was especially true for the Introduction, Background and Methods but did not occur for the Conclusions. The highly cited articles that had the greatest share of citations for each section except the Conclusions were investigated to find out the cause of the citing section unevenness (see Appendix). In the Methods section clearly gave section specific information for the two highly cited articles. The articles with most citations from the Introduction, Background or Discussion due to structural reasons rather than for a distinctive contribution. For example, the articles with mostly Introduction citations did not obviously provide a context-setting role but were cited by articles lacking a separate literature review.

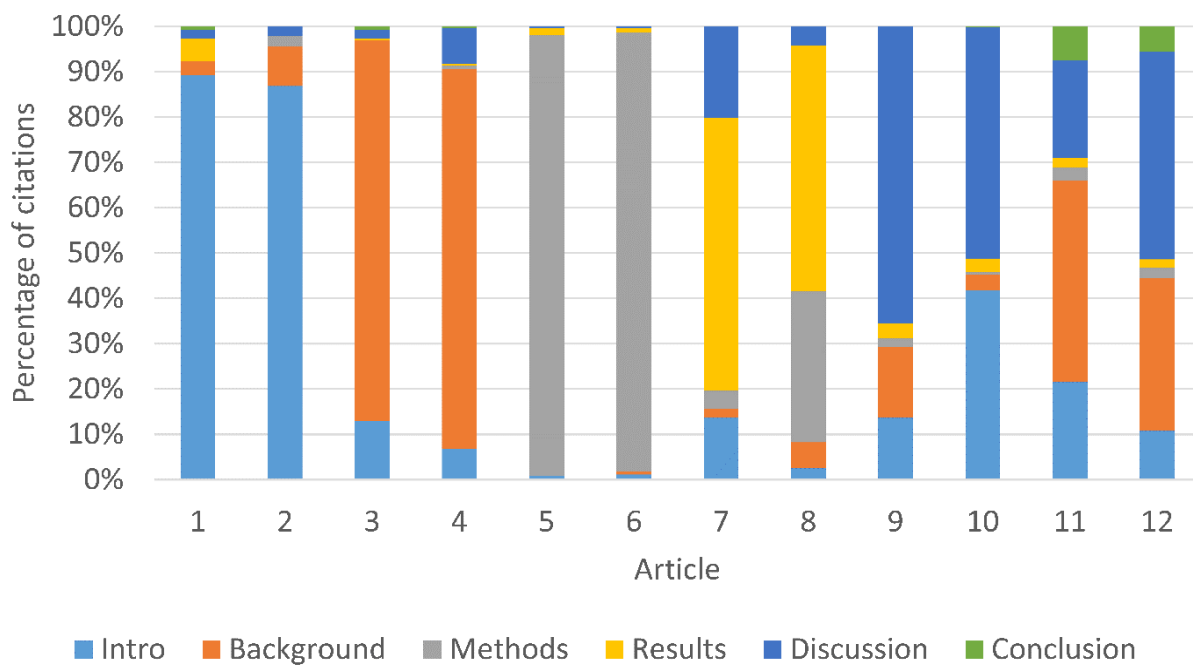

Figure 9. Shares of citation for the articles with at least 100 citations from the six sections and the highest share in one of the six sections (two per section). See Appendix 2 for article information.

## 5 Discussion

The results show that whilst there are disciplinary differences, in most fields articles host substantial numbers of citations from each of the six sections investigated, except for the Conclusions and, to a lesser extent, the Results (Figure 1). The Introduction is the most common source of citations, but this is no longer true when focusing on cited articles with DOIs (Figure 2). From the perspective of these cited articles, five of the six sections (except the Conclusions) could host substantial numbers of citations. The combination of sections sending citations varies greatly by field, with no obvious commonalities between similar fields. The extent to which information about the section citing an article provides distinctive evidence about the context of the citation depends on whether articles are typically cited by multiple sections. For the 21 cited fields and six sections analysed in detail, there were 23 cases where a section hosted relatively unique citations (unlabelled graphs in Figures 3 to 7). This was most common for the Introduction, followed by the Methods, Background and Discussion. There did not seem to be a broad disciplinary pattern in the results because related fields often displayed different patterns. From a different perspective, if citations are counted from only one of the six sections then the results have a low correlation with total citations or with counts of citations from any other section (Table 1). The low correlations tend to contradict the other results by suggesting that citations from each section provide different information, but this may be primarily due to most articles having few citations. Finally, the existence of highly cited articles with citations primarily from only one of the first five sections (Figure 9) confirms that these sections can give distinctive citation context information in some cases even if the other evidence (Figures 3 to 8) suggests that this is not common.

**RQ1**: The information about sections hosting citations (Figure 1) extends two previous studies of single information science journals that found the Introduction (Hu, Chen, & Liu, 2013) and Literature Review sections to contain the most citations (Ding, Liu, Guo, & Cronin, 2013). The results also extend a prior study of seven PLOS journals which found that the Introduction contained the most references (Bertin, Atanassova, Gingras, & Larivière, 2016); this was true for about half of the fields analysed here. The fact that the Discussion section, which is near the end, can contain the most references in some fields agrees with a prior study of the PMC Open Access collection that found the peak position for references in some fields was about two thirds of the length of the article (Boyack, van Eck, Colavizza, & Waltman, 2018). Nevertheless, the current results suggest greater heterogeneity between fields in the sections containing many citations, such the five Enabling & Strategic Technologies sections. When the results are restricted to citations with DOIs and analysed from the perspective of the field of the cited article, even more heterogeneity is present (Figure 2).

**RQ2**: The results reveal substantial disciplinary differences in the patterns of citations between sections: the extent to which articles cited by one section are also cited by another. This issue that has not previously been investigated. Most importantly, these results suggest that overlaps between sections in the articles cited is normal for most fields and sections. Whilst there are exceptions (mainly for Introduction and Methods citations), this suggests that counting citations separately by the six sections reported here may not give sharply differentiated citation context information. It is more likely to give fuzzier information with articles attracting citations from two or more sections even if the underlying reason for citing is the same. This applies whether considering citation counts for individual highly cited articles or for large collections of articles.

**RQ3**: The weak, and sometimes negative, correlations between citation counts from one section with citation counts from other sections also addresses an issue that has not previously been investigated. The low correlations are probably due to large numbers of articles with only one citation, however, and so the magnitude of the correlations is not informative.

**RQ4**: The (unsurprising) existence of highly cited papers that predominantly attract citations from one article section is another novel type of finding, although it has been shown previously that highly cited papers can be mainly selected for methods purposes (Small, 2018). This is now shown to be true also for the Introduction, Background, Discussion and, to a lesser extent, the Results. It was not true for the Conclusions in the current dataset.

The lack of uniqueness for citations from most sections/fields can be due to several reasons. First, there is a natural overlap between the function of the Introduction and Background and this is exacerbated if some articles in a field omit the background section and include a literature review within the introduction. Second, it is natural to return to background material introduced earlier in the Discussion, giving an overlap in citations between the Introduction, Background and Discussion. This is evident in Figures 3 and 4 and, to a lesser extent, Figure 7. The Results section often contains few citations but may include an element of initial results discussion and hence could overlap with Discussion citations (and hence Introduction and Background citations). As mentioned above, Methods citations can overlap with other sections due to methods-oriented papers or perhaps, in the case of key methods, the need to refer to them when introducing a paper or discussing the results.

In addition to the above considerations, the structure of an article can affect the section in which references are placed. For example, if the Methods is at the end then methodological references might be included in the Results. The extent to which a literature review is regarded as essential to understand a contribution may also affect where references occur (e.g., background literature might be cited in the Discussion rather than the Introduction, omitting any Background section, if early background is believed to be unnecessary).

This research is limited by using a biased subset of the scientific literature and articles from multiple years without a fixed citation window for all analyses except the correlations. This complicates interpretation of some of the findings. They are also limited by tracking articles by DOI, thereby ignoring citations omitting DOIs and citations to articles lacking DOIs. For RQ2, the inclusion of only 32.3% of citations also affects the shape of the graphs because an unknown proportion of citations to each section will be lost due to missing DOIs. The strength of the conclusions about individual fields is also affected by the indirect method used to investigate whether sections cite different articles (Figures 3 to 8), leading to the use of heuristics to judge whether two sections overlap. The answer to RQ4 is not addressed for each field and so may not be true for all disciplines.

# 6 Conclusions

An important practical implication of the research is that section headings are not reliable as indicators of the purpose of a citation, in terms of the six-section structure used here. Given the prior evidence that some citations can be low value – and particularly those in the Introduction and Background (Maričić, Spaventi, Pavičić, & Pifat-Mrzljak, 1998; Tang & Safer, 2008), the difficulty with using section headings to separate citations in most cases is unfortunate. There are field differences in the uses of different sections and journal or field conventions can push citations into sections where they would not belong in other fields. A more fine-grained approach may be needed that either does not rely on section headings alone to give more precise results.

Despite the above negative conclusions, citation counting by section may still be useful when applied to aggregate data or highly cited articles, when anomalies might be expected to cancel out (as is standard for citation analysis (van Raan, 1998). For example, if highly cited article A had a higher proportion of Introduction citations than highly cited article B from the same field then this could be taken as weak evidence that article A was more useful for the introductory setup of new research than for its substantive content. Expert judgement (as in the case of citations) would be needed to decide whether any difference was meaningful because of the possibility that article structure differences in key journals is the root cause. Comparisons between fields would not be reasonable because of differing article structures. Counting citations by section would be most useful to help identify articles with context setting (mainly Introduction citations) and methodological (mainly Methods citations). There is evidence for the former in the relatively unique nature of Introduction citations (Figure 3) as well as prior research suggesting that perfunctory citations are most likely to occur in the Introduction. There is evidence for the latter in the relatively unique nature of Methods citations in some fields (Figure 5), prior research suggesting that many highly cited papers primarily attract methods-related citations, and the two highly cited papers investigated here for containing primarily Methods section citations. Following this logic, the remaining articles that do not primarily attract Introduction or Methods citations might be formed into a third group.

# 8   Appendix 1

Tables A1, A2, A3 and A4 give supporting information for the main results.

Table A1. Sample sizes (citing articles) for Figures 3-8. Articles are included if they have a citation count of at least 1 in any of the sections.

| Subject | Short name | All | Intro | Backgr | Meth | Results | Discuss | Conc |
|---|---|---|---|---|---|---|---|---|
| Agriculture, Fisheries & Forestry | Agriculture | 12309 | 3585 | 4523 | 521 | 337 | 6267 | 6 |
| Biology | Biology | 42512 | 15229 | 16483 | 1650 | 5273 | 22755 | 189 |
| Biomedical Research | Biomedical | 183660 | 59298 | 69129 | 7696 | 31537 | 100146 | 718 |
| Built Environment & Design | Built Env | 181 | 81 | 61 | 10 | 6 | 54 | 0 |
| Chemistry | Chemistry | 17331 | 11017 | 4013 | 588 | 2099 | 2756 | 26 |
| Clinical Medicine | Clinical Med | 341544 | 107559 | 123361 | 17330 | 20356 | 214880 | 781 |
| Communication & Textual Studies | Commun | 102 | 51 | 25 | 15 | 2 | 31 | 0 |
| Earth & Environmental Sciences | Earth & Env | 5971 | 3631 | 690 | 173 | 712 | 1701 | 8 |
| Economics & Business | Econ & Bus | 1577 | 546 | 649 | 275 | 63 | 578 | 11 |
| Enabling & Strategic Technologies | En & Strateg | 36313 | 15763 | 16819 | 3368 | 7672 | 9848 | 335 |
| Engineering | Engineering | 5295 | 2818 | 1219 | 330 | 283 | 1720 | 3 |
| General Arts, Humanities & Social Sciences | Art Hum Soc | 189 | 32 | 94 | 6 | 2 | 105 | 2 |
| General Science & Technology | Sci & Tech | 51256 | 23412 | 16171 | 2239 | 9733 | 27872 | 382 |
| Historical Studies | History | 394 | 155 | 106 | 21 | 23 | 173 | 0 |
| Information & Communication Technologies | ICTs | 4136 | 1476 | 2137 | 491 | 236 | 1590 | 20 |
| Mathematics & Statistics | Maths | 2723 | 618 | 981 | 1381 | 296 | 659 | 21 |
| Philosophy & Theology | Phil & Theol | 530 | 109 | 276 | 17 | 13 | 259 | 2 |
| Physics & Astronomy | Phys & Astro | 12960 | 9918 | 1010 | 494 | 2353 | 1159 | 26 |
| Psychology & Cognitive Sciences | Psychology | 24255 | 13590 | 5656 | 2323 | 562 | 10447 | 49 |
| Public Health & Health Services | Public Health | 52098 | 11165 | 28045 | 6886 | 1837 | 29255 | 234 |
| Social Sciences | Social Sci | 3709 | 1296 | 1317 | 713 | 188 | 1370 | 14 |
| Visual & Performing Arts | Arts | 10 | 9 | 1 | 0 | 0 | 0 | 0 |
| **Total** | | **799055** | **281358** | **292766** | **46527** | **83583** | **433625** | **2827** |

Table A2. Citations with DOIs by originating section and field. Fractional counting is used when a paper is cited by multiple sections. Total number of citations extracted: 30,496,604.

| Field | Other | Intro | Backgr | Meth | Results | Discuss | Conc |
|---|---|---|---|---|---|---|---|
| Agriculture | 129952 | 93752 | 9960 | 9569 | 8787 | 134060 | 1053 |
| Biology | 461771 | 432279 | 33068 | 40625 | 75536 | 424279 | 8516 |
| Biomedical | 1978549 | 1617572 | 126697 | 139258 | 508245 | 1496700 | 40230 |
| Built Env | 2088 | 404 | 3 | 17 | 18 | 60 | 21 |
| Chemistry | 564528 | 156215 | 79845 | 4568 | 7796 | 10825 | 5611 |
| Clinical Med | 1739641 | 1348880 | 323613 | 336628 | 256169 | 2724974 | 36088 |
| Commun | 1750 | 101 | 8 | 26 | 10 | 21 | 22 |
| Earth & Env | 40716 | 22438 | 1007 | 1510 | 2138 | 5708 | 654 |
| Econ & Bus | 12311 | 2166 | 130 | 280 | 178 | 243 | 147 |
| En & Strateg | 639597 | 158819 | 223024 | 126600 | 123390 | 161738 | 22180 |
| Engineering | 21079 | 9317 | 7613 | 5846 | 1645 | 9065 | 472 |
| Art Hum Soc | 5975 | 216 | 6 | 19 | 8 | 48 | 44 |
| Sci & Tech | 5272527 | 3450490 | 14726 | 107684 | 558668 | 971303 | 352197 |
| History | 15782 | 460 | 3 | 13 | 40 | 189 | 66 |
| ICTs | 26797 | 25050 | 18101 | 25100 | 5122 | 56336 | 864 |
| Maths | 14502 | 7005 | 44 | 333 | 126 | 1105 | 99 |
| Phil & Theol | 12009 | 1831 | 2894 | 881 | 525 | 3669 | 409 |
| Phys & Astro | 152662 | 244812 | 58 | 1568 | 11549 | 46554 | 2363 |
| Psychology | 273497 | 436613 | 3767 | 16058 | 17995 | 217725 | 4456 |
| Public Health | 181240 | 73702 | 222205 | 159988 | 25741 | 414313 | 8932 |
| Social Sci | 40535 | 11250 | 335 | 1158 | 685 | 2245 | 678 |
| Arts | 218 | 20 | 0 | 8 | 3 | 5 | 0 |
| Total | 11587726 | 8093391 | 1067107 | 977738 | 1604374 | 6681164 | 485103 |
| Total % | 38% | 27% | 3% | 3% | 5% | 22% | 2% |

Table A3. Median correlation between citation counts calculated only from the named section (n=22 fields; median sample size: 9245.5).

| Section | Back | Meth | Res | Disc | Conc | **All** |
|---------|------|------|------|------|------|---------|
| Introduction | -0.15 | -0.13 | -0.03 | -0.13 | 0.01 | 0.34 |
| Background | 1.00 | 0.03 | -0.05 | -0.02 | 0.04 | 0.38 |
| Methods | | 1.00 | 0.08 | -0.02 | 0.02 | 0.17 |
| Results | | | 1.00 | -0.01 | 0.04 | 0.17 |
| Discussion | | | | 1.00 | 0.02 | 0.37 |
| Conclusions | | | | | 1.00 | 0.08 |

Table A4. Number of fields with a positive correlation between citation counts calculated only from the named section (n=22 fields). High and low percentages are bold.

| Section | Back | Meth | Res | Disc | Conc | All |
|---------|------|------|-----|------|------|-----|
| Introduction | 5% | **9%** | 36% | **9%** | 59% | 100% |
| Background | 100% | 59% | 27% | 45% | 77% | 100% |
| Methods | | 100% | **91%** | 36% | 77% | 100% |
| Results | | | 100% | 41% | **91%** | 95% |
| Discussion | | | | 100% | 64% | 95% |
| Conclusions | | | | | 100% | 91% |

# 9 Appendix 2: Section-specific sources of citations to highly cited articles

Most (116 out of 118) of the **Introduction** citations to article 1 (Table A5) were from the journal *Nature Communications*, which requires authors to put the literature review in the introduction section ("The main text of an Article should begin with an introduction (without heading) of referenced text that expands on the background of the work (some overlap with the abstract is acceptable), followed by sections headed Results, Discussion (if appropriate) and Methods (if appropriate).", https://www.nature.com/ncomms/submit/article). This article is often first in the reference list (53 times out of 116 in the journal) and is a perfunctory citation (e.g., 10.1038/s41467-017-01108-z, "The discovery of topological insulators has ignited great research interest in the novel physical properties of topological materials in the past decade [refs]"). Article 2 was also the first cited in 52 of its 118 Introduction citations (the identical number of Introduction citations is a coincidence with article 1). These are probably relatively perfunctory (e.g., 10.1364/BOE.2.002690, "Optical Coherence Tomography (OCT) is a non-invasive, depth resolved, medical imaging modality, which is well suited as a tool for diagnostic visualization of the retinal structures in-vivo [ref]."). About half (52 of the 118 Introduction citations were from *Biomedical Optics Express*).

The largest share (37 out of 123) of the **Background** citations to article 3 are from *BMC Bioinformatics* and in 51 cases it is the first citation, again suggesting a relatively perfunctory role. This journal mandates a structure of Background-Results-Conclusions (https://bmcbioinformatics.biomedcentral.com/submission-guidelines/preparing-your-manuscript/research-article) and so the Background section plays the role of the Introduction and Background sections combined. These citations again seem to be perfunctory (e.g., "Microarrays are well known for their success in studying gene expression [ref]" starts 10.1186/1471-2105-10-293). Article 4 is usually the first in reference lists if it is mentioned in Background sections (111 out of 153) and is mostly cited by the *Malaria*

*Journal* (113 out of 153), which mandates a Background section instead of an Introduction (Background-Methods-Results-Conclusions-Trial registration: https://malariajournal.biomedcentral.com/submission-guidelines/preparing-your-manuscript/research-article). It may be cited as a relatively perfunctory context-setting role (e.g., "close to 40% of the world's population live in countries where the disease is endemic and nearly 247 million people suffer from the disease every year [ref]", 10.1186/1475-2875-10-274).

The largest share (68 out of 276) of the **Methods** citations to article 5 were from *Nature Scientific Reports*, where it was first on the reference list only 6 times. It seems to be cited to acknowledge software use (e.g., 10.1038/s41598-018-24009-7, "a likelihood ratio test of LD implemented with the ARLEQUIN software[ref]"). The largest share (112 out of 327) for article 6 was also from *Nature Scientific Reports*, to acknowledge using the UCHIME algorithm ("and filtered for length and putative chimeric PCR products (UCHIME)", 10.1093/bioinformatics/btr381).

The largest share (29 out of 76) of the **Results** citations to article 7 were from *Cell Disease & Death*. This reference was used to give context ("Autophagy is a highly regulated process, in which the activities of autophagy-related (ATG) proteins are involved [ref]. Next, we determined whether BBM affects the activities of autophagy-related proteins in breast cancer cells.", 10.1038/s41419-018-0276-8) and ("Moreover, rapamycin protection against PrP90-231 neurotoxicity was abolished by autophagy inhibition induced by 3-methyladenine (3-MA)[ref] (Fig. 8a).", 10.1038/s41419-017-0252-8). These articles have a Materials and Methods section at the end of the article and seem to include additional methods details within the results section to guide the reader because of this. Thus, the Results citations are methodological citations. A third (33 out of 96) of the Results citations to article 8 were from *Nature Scientific Reports*, which also has its methods section at the end. These references also seemed to play a methodological role (e.g., "In order to confirm our data concerning the Wnt pathway, we proceeded to an additional KEGG Automatic Annotation[ref]", 10.1038/s41598-017-15557-5).

The sources of citations in the **Discussion** citations to article 9 varied (no more than 5 [*Nature Scientific Reports*] out of 74 from a single journal). These citations seemed to appropriately contextualise the results, but sometimes with an indirect role that might be more appropriate for a background section (e.g., "Systemic inflammation is also known to induce direct changes on HDL particle and APOA1 molecule concentrations. For example, serum amyloid A (SAA) expression is markedly increased in response to acute and chronic inflammation[ref]", 10.1038/s41598-017-05415-9). Thus, high levels of citation in the Discussion section might be indicative of citations from articles with short introductions, so that additional background is needed in the Discussion. The sources of citations in the Discussion to article 10 were also varied (no more than 13 [*Nature Scientific Reports*] out of 67 from a single journal), and again plays an appropriate backgrounding discussion role (e.g., "The authors interpreted the significantly right lateralized putamen activity during non-auditory counting as an indication of this region being involved in sustained attention tasks requiring working memory. Right hemisphere dominance during attention as well as specific auditory attention tasks has been suggested repeatedly in literature[ref]", 10.1038/s41598-017-08728-x).

The two **Conclusions** articles are not discussed since these attracted too few Conclusions references to draw inferences from.

Table A5. Two articles with the highest share of citations from each section: qualification: at least 100 citations from the six sections combined. The percentage of citations to the article that were not attributed to one of the six sections is reported at the end. See Figure 9 for citation shares by section for these articles.

| Art. | Journal | Year | Title | Comment | Other secs |
|------|---------|------|-------|---------|------------|
| I 1 | Reviews of Modern Physics | 2010 | Colloquium: Topological insulators | Opens new avenue for research in quantum physics | 13% |
| I 2 | Science | 1991 | Optical coherence tomography | Introduces optical coherence tomography method for biological systems. | 45% |
| B 3 | Science | 1995 | Quantitative monitoring of gene expression patterns with a complementary DNA microarray | Method to detect gene expressions | 16% |
| B 4 | Nature | 2005 | The global distribution of clinical episodes of Plasmodium falciparum malaria | Method to detect global distribution of Malaria and Malaria prevalence data from 2002 | 16% |
| M 5 | Molecular Ecology Resources | 2010 | Arlequin suite ver 3.5: a new series of programs to perform population genetics analyses under Linux and Windows | Software to aid experiments. | 74% |
| M 6 | Bioinformatics | 2011 | UCHIME improves sensitivity and speed of chimera detection | Proof of algorithm working for detecting biological sequences. | 86% |
| R 7 | Autophagy | 2016 | Guidelines for the use and interpretation of assays for monitoring autophagy (3rd edition) | Guidelines for interpreting cell destruction experiment results | 59% |
| R 8 | Nucleic Acids Research | 2016 | KEGG as a reference resource for gene and protein annotation | "biological interpretation of genome sequences" database. | 60% |
| D 9 | NEJM | 1999 | Acute-phase proteins and other systemic responses to inflammation | Review of research into systemic responses to inflammation. | 39% |
| D 10 | Nature Reviews Neuroscience | 2002 | Control of goal-directed and stimulus-driven attention in the brain | "review[s] evidence for partially segregated networks of brain areas that carry out different attentional functions" | 73% |
| C 11 | The Lancet | 2010 | Health professionals for a new century: Transforming education to strengthen health systems in an interdependent world | NA: No article had a high share of conclusions references. | 38% |
| C 12 | The Lancet | 2005 | How can we achieve and maintain high-quality performance of health workers in low-resource settings? | NA: No article had a high share of conclusions references. | 34% |